%% file: fenwick.tex
\documentclass{llncs}

\usepackage{mathtools}
\usepackage{makeidx}  
\usepackage[sort,nocompress]{cite}
\usepackage[utf8]{inputenc}

\newcommand{\bigO}{\mathcal{O}}

\begin{document}

\title{Succinct Partial Sums and Fenwick Trees}
%
%
\author{Philip Bille \and Anders Roy Christiansen \and Nicola Prezza  \and  Frederik Rye Skjoldjensen}
\authorrunning{} 
%
%
\institute{Technical University of Denmark, DTU Compute, Kgs. Lyngby, Denmark\\
\email{\{phbi,aroy,npre,fskj\}@dtu.dk}}
\maketitle

\input{abstract}
\input{introduction}
\input{datastructure}

\bibliographystyle{splncs03}
\bibliography{fenwick}

\end{document}

%% file: abstract.tex

\begin{abstract}
	We consider the well-studied \emph{partial sums} problem in succint space where one is to maintain an array of $n$ $k$-bit integers subject to updates such that partial sums queries can be efficiently answered. We present two succint versions of the Fenwick Tree -- which is known for its simplicity and practicality.
	Our results hold in the encoding model where one is allowed to reuse the space from the input data. Our main result is the first that only requires $nk + o(n)$ bits of space while still supporting sum/update in $\bigO(\log_bn)$ / $\bigO(b \log_bn)$ time where $2 \leq b \leq \log^{\bigO(1)}n$. The second result shows how optimal time for sum/update can be achieved while only slightly increasing the space usage to $nk + o(nk)$ bits.
	Beyond Fenwick Trees, the results are primarily based on bit-packing and sampling -- making them very practical -- and they also allow for simple optimal parallelization.

	\keywords{Partial sums, Fenwick tree, succinct, parallel}
\end{abstract}

%% file: introduction.tex

\section{Introduction}
Let $A$ be an array of $k$-bits integers, with $|A| = n$. The \emph{partial sums} problem is to build a data structure maintaining $A$ under the following operations.  

\begin{itemize}
\item \texttt{sum}$(i)$: return the value $\sum_{t=1}^{i}A[t]$.
\item \texttt{search}$(j)$: return the smallest $i$ such that \texttt{sum}$(i) \geq j$.
\item \texttt{update}$(i, \Delta)$: set $A[i] \leftarrow A[i] + \Delta$, for some  $\Delta$ such that $0 \leq A[i] + \Delta < 2^k$.
\item \texttt{access}$(i)$: return $A[i]$. 
\end{itemize}
Note that \texttt{access}$(i)$ can implemented as \texttt{sum}$(i) - $\texttt{sum}$(i-1)$ and we therefore often do not mention it explicitly.  

The partial sums problem is one the most well-studied data structure problems~\cite{fredman1982complexity, yao1985complexity, fredman1989cell, dietz1989optimal, fenwick1994new, raman2001succinct, hon2011succinct, patrascu2006logarithmic}. In this paper, we consider solutions to the partial sums problem that are \emph{succinct}, that is, we are interested in data structures  that use space close to the information-theoretic lower bound of $nk$ bits. We distinguish between  \emph{encoding data structures} and \emph{indexing data structures}. Indexing data structures are required to store the input array $A$ verbatim along with additional information to support the queries, whereas encoding data structures have to support operations without consulting the input array. 


In the indexing model Raman et al.~\cite{raman2001succinct} gave a data structure that supports \texttt{sum}, \texttt{update}, and \texttt{search} in $\bigO(\log n /\log \log n)$ time while using $nk + o(nk)$ bits of space. This was improved and generalized by Hon et al.~\cite{hon2011succinct}. Both of these papers have the constrain $\Delta \leq \log^{\bigO(1)}n$.
The above time complexity is nearly optimal by a lower bound of Patrascu and Demaine~\cite{patrascu2006logarithmic} who showed that \texttt{sum}, \texttt{search}, and \texttt{update} operations takes $\Theta(\log_{w/\delta} n)$ time per operation, where $w \geq \log n$ is the word size and $\delta$ is the number of bits needed to represent $\Delta$. In particular, whenever $\Delta = \log^{\bigO(1)}n$ this bound matches the $\bigO(\log n /\log \log n)$ bound of Raman et al.~\cite{raman2001succinct}. 

Fenwick~\cite{fenwick1994new} presented a simple, elegant, and very practical encoding data structure. The idea is to replace entries in the input array $A$ with partial sums that cover $A$ in an implicit complete binary tree structure. The operations are then implemented by accessing at most $\log n$ entries in the array. The Fenwick tree uses $nk + n\log n$ bits and supports all operations in $O(\log n)$ time.
In this paper we show two succinct $b$-ary versions of the Fenwick tree. In the first version we reduce the size of the Fenwick tree while improving the \texttt{sum} and \texttt{update} time. In the second version we obtain optimal times for \texttt{sum} and \texttt{update} without using more space than the previous best succinct solutions~\cite{raman2001succinct,hon2011succinct}. All results in this paper are in the RAM model.

\paragraph{Our results}
We show two encoding data structures that gives the following results.

\begin{theorem}\label{th1} 
	We can replace $A$ with a succinct Fenwick tree  of $nk+o(n)$ bits supporting \texttt{sum}, \texttt{update}, and \texttt{search} queries in $\bigO(\log_bn)$, $\bigO(b\log_bn)$, and $\bigO(\log n)$ time, respectively, for any $2 \leq b \leq \log^{\bigO(1)}n$.
\end{theorem}

\begin{theorem}\label{th2} 
	We can replace $A$ with a succinct Fenwick tree of $nk  + o(nk)$ bits supporting \texttt{sum} and \texttt{update} queries in optimal $\bigO(\log_{w/\delta} n)$ time and \texttt{search} queries in  $\bigO(\log n)$ time. 
\end{theorem}


%% file: datastructure.tex

\section{Data structure}

For simplicity, assume that $n$ is a power of $2$. The Fenwick tree is an implicit data structure replacing a word-array $A[1,\dots,n]$ as follows:

\begin{definition} Fenwick tree of $A$~\cite{fenwick1994new}. 
If $n=1$, then leave $A$ unchanged. Otherwise, 
divide $A$ in consecutive non-overlapping blocks of two elements each and replace the second element $A[2i]$ of each block with $A[2i-1]+A[2i]$, for $i=1,\dots,n/2$. Then, recurse on the sub-array $A[2,4,\dots,2i,\dots, n]$. 
\end{definition}

To answer $sum(i)$, the idea is to write $i$ in binary as $i = 2^{j_1}+2^{j_2}+\dots+2^{j_k}$ for some $j_1>j_2>\dots>j_k$. Then there are $k\leq\log n$ entries in the Fenwick tree, that can be easily computed from $i$, whose values added together yield $sum(i)$.
In Section \ref{sec:b-ary} we describe in detail how to perform such accesses.
As per the above definition, the Fenwick tree is an array of $n$ indices. If represented compactly, this array can be stored in $nk+n\log n$ bits. In this section we present a generalization of Fenwick trees taking only succinct space. 

\subsection{Layered b-ary structure}\label{sec:b-ary}

We first observe that it is easy to generalize Fenwick trees to be $b$-ary, for $b\geq 2$: we divide $A$ in blocks of $b$ integers each, replace the first $b-1$ elements in each block with their partial sum, and fill the remaining $n/b$ entries of $A$ by recursing on the array $A'$ of size $n/b$ that stores the sums of each block. This generalization gives an array of $n$ indices supporting \texttt{sum}, \texttt{update}, and \texttt{search} queries on the original array in $\bigO(\log_bn)$, $\bigO(b\log_bn)$, and $\bigO(\log n)$ time, respectively. 
We now show how to reduce the space of this array. 

Let $\ell = \log_b n$.
We represent our $b$-ary Fenwick tree $T_b(A)$ using $\ell+1$ arrays (layers) $T_b^1(A), \dots, T_b^{\ell+1}(A)$. For simplicity, we assume that $n=b^e$ for some $e\geq 0$ (the general case is then straightforward to derive).  To improve readability, we define our layered structure for the special case $b=2$, and then sketch how to extend it to the general case $b\geq 2$. 
Our layered structure is defined as follows. If $n=1$, then $T_2^1(A)=A$. Otherwise:
\begin{itemize}
	\setlength\itemsep{5pt}
	\item $T_2^{\ell+1}(A)[i] = A[(i-1)\cdot 2+1]$, for all $i=1,\dots, n/2$. Note that $T_2^{\ell+1}(A)$ contains $n/2$ elements.
	\item Divide $A$ in blocks of $2$ elements each, and build an array $A'[j]$ containing the $n/2$ sums of each block, i.e. $A'[j] = A[(j-1)\cdot 2 + 1] + A[(j-1)\cdot 2 + 2]$, for $j=1,\dots, n/2$. Then, the next layers are recursively defined as $T_2^{\ell}(A) \leftarrow T_2^{\ell}(A'), \dots, T_2^{1}(A) \leftarrow T_2^{1}(A')$.
\end{itemize}

\begin{figure}
	\begin{center}
		\includegraphics[scale=0.8]{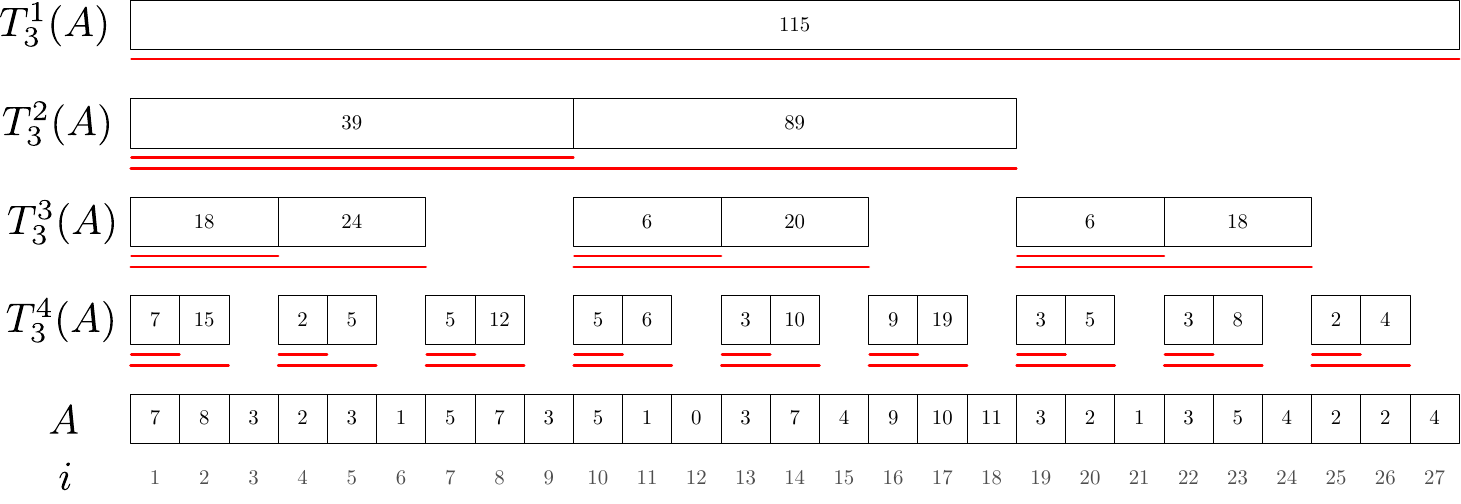}
	\end{center}
	\caption{Example of our layered structure with $n=27$ and $b=3$. Horizontal red lines show the portion of $A$ covered by each element in $T_3^j(A)$, for $j=1,\dots,\log_bn+1$. To access the $i$-th partial sum, we proceed as follows. Let, for example, $i = 19 = (0201)_3$. The only nonzero digits in $i$ are the $2$-nd and $4$-th most significant. This gives us $o_2 = 2\cdot(0)_3 + 2 = 2$ and $o_4 = 2\cdot (020)_3 + 1 = 13$. Then, $sum(19) = \mathcal T_3^2(A)[2] + T_3^4(A)[13] = 89 + 3 = 92$.}\label{fenwickfig}
\end{figure}

For general $b\geq 2$, $T_b^{\ell+1}(A)$ is an array of $\frac{n(b-1)}{b}$ elements that stores the $b-1$ partial sums of each block of $b$ consecutive elements in $A$, while $A'$ is an array of size $n/b$ containing the complete sums of each block. 
In Figure \ref{fenwickfig} we report an example of our layered structure with $b=3$.
It follows that elements of $T_b^i(A)$, for $i>1$, take at most $k+(\ell-i+2)\log b$ bits each.
Note that arrays  $T_b^1(A), \dots, T_b^{\ell+1}(A)$ can easily be  packed contiguously in a word array while preserving constant-time access to each of them. This saves us $\bigO(\ell)$ words that would otherwise be needed to store pointers to the arrays. 
Let $S_b(n,k)$ be the space (in bits) taken by our layered structure. This function satisfies the recurrence
$$
\begin{array}{l}
S_b(1,k) = k\\
S_b(n,k) = \frac{n(b-1)}{b}\cdot \left(k+\log b\right) + S_b(n/b,k+\log b)
\end{array}
$$
Which unfolds to
$
S_b(n,k) = \sum_{i=1}^{\log_bn+1} \frac{n(b-1)}{b^i}\cdot \left(k+i\log b\right).
$
Using the identities $\sum_{i=1}^{\infty} 1/b^i = 1/(b-1)$ and $\sum_{i=1}^{\infty} i/b^i = b/(b-1)^2$, one can easily derive that $S_b(n,k) \leq nk + 2n\log b$.

We now show how to obtain the time bounds stated in Theorem \ref{th1}. In the next section, we reduce the space of the structure without affecting query times.

\paragraph{Answering sum} 
Let the notation $(x_1x_2\dots x_t)_b$, with $0\leq x_i <b$ for $i=1,\dots,t$, represent the number $\sum_{i=1}^{t}b^{t-i}x_i$ in base $b$.
$sum(i)$ queries on our structure are a generalization (in base $b$) of $sum(i)$ queries on standard Fenwick trees.
Consider the base-$b$ representation $x_1x_2\dots x_{\ell+1}$ of $i$, i.e. $i = (x_1x_2\dots x_{\ell+1})_b$ (note that we have at most $\ell+1$ digits since we enumerate indexes starting from 1). Consider now all the positions $1\leq i_1<i_2 <\dots < i_t \leq \ell+1$ such that $x_j\neq 0$, for $j = i_1, \dots, i_t$. The idea is that each of these positions $j = i_1, \dots, i_t$ can be used to compute an offset $o_j$ in $T^{j}_b(A)$. Then, $sum(i) = \sum_{j = i_1, \dots, i_t} T^{j}_b(A)[o_j]$. 
The offset $o_j$ relative to the $j$-th most significant (nonzero) digit of $i$ is defined as follows. 
If $j=1$, then $o_j = x_1$. Otherwise,  $o_j = (b-1)\cdot (x_1\dots x_{j-1})_b + x_j$. Note that we scale by a factor of $b-1$ (and not $b$) as the first term in this formula as each level $T^{j}(A)$ stores only $b-1$ out of $b$ partial sums (the remaining sums are passed to level $j-1$). Note moreover that each $o_j$ can be easily computed in constant time and \emph{independently from the other offsets} with the aid of modular arithmetic. It follows that \texttt{sum} queries are answered in $\bigO(\log_bn)$ time. 
See Figure \ref{fenwickfig} for a concrete example of \texttt{sum}.

\paragraph{Answering update} The idea for performing $update(i,\Delta)$ is analogous to that of $sum(i)$. We access all levels that contain a partial sum covering position $i$ and update at most $b-1$ sums per level. Using the same notation as above, for each $j = i_1, \dots, i_t$ such that $x_j\neq 0$, we update $T^{j}_b(A)[o_j + l] \leftarrow T^{j}_b(A)[o_j + l]+ \Delta$ for $l = 0, \dots, b-x_j-1$. This procedure takes $\bigO(b\log_bn)$ time. 

\paragraph{Answering search} To answer $search(j)$ we start from $T^{1}_b(A)$ and simply perform a top-down traversal of the implicit B-tree of degree $b$ defined by the layered structure. At each level, we perform $\bigO(\log b)$ steps of binary search to find the new offset in the next level. There are $\log_bn$ levels, so \texttt{search} takes overall $\bigO(\log n)$ time. 

\subsection{Sampling}\label{sec:sampling}


Let $0 < d \leq n$ be a sample rate, where for simplicity we assume that $d$ divides $n$. Given our input array $A$, we derive an array $A'$ of $n/d$ elements containing the sums of groups of $d$ adjacent elements in $A$, i.e. $A'[i] = \sum_{j=1}^{d} A[(i-1)\cdot d + j]$, $i=1, \dots, d$. We then compact $A$ by removing $A[j\cdot d]$ for $j=1, \dots, n/d$, and by packing the remaining integers in at most $nk(1-1/d)$ bits. We build our layered $b$-ary Fenwick tree $T_b(A')$ over $A'$.  It is clear that queries on $A$ can be solved with a query on $T_b(A')$ followed by at most $d$ accesses on (the compacted) $A$. The space of the resulting data structure is $nk(1 - 1/d) + S_b(n/d, k+\log d) \leq nk + \frac{n\log d}{d} + \frac{2n\log b}{d}$ bits. In order to retain the same query times of our basic layered structure, we choose $d = (1/\epsilon)\log_bn$ for any constant $\epsilon>0$ and obtain a space occupancy of
$
nk + \epsilon \left(\frac{n\log\log_bn}{\log_bn} + \frac{2n\log b}{\log_bn}\right)
$
bits. For $b \leq \log^{\bigO(1)}n$, this space is $nk + o(n)$ bits. Note that---as opposed to existing succinct solutions---the low-order term does not depend on $k$.

\section{Optimal-time \texttt{sum} and \texttt{update}}

In this section we show how to obtain optimal running times for \texttt{sum} and \texttt{update} queries in the RAM model. 
We can directly apply the word-packing techniques described in \cite{patrascu2006logarithmic} to speed-up queries; here we only sketch this strategy, see~\cite{patrascu2006logarithmic} for full details. Let us describe the idea on the structure of Section \ref{sec:b-ary}, and then plug in sampling to reduce space usage. We divide arrays $T^{j}_b(A)$ in blocks of $b-1$ entries, and store one word ($w$ bits) for each such block. We can  pack $b-1$ integers  of at most $w/(b-1)$ bits each (for an opportune $b$, read below) in the word associated with each block. Since blocks of $b-1$ integers fit in a single word, we can easily answer \texttt{sum} and \texttt{update} queries on them in constant time. 
\texttt{sum} queries on our overall structure can be answered as described in Section \ref{sec:b-ary}, except that now we also need to access one of the packed integers at each level $j$ to correct the value read from $T^{j}_b(A)$.
To answer \texttt{update} queries, the idea is to perform \texttt{update} operations on the packed blocks of integers in constant time exploiting bit-parallelism instead of updating at most $b-1$ values of $T^{j}_b(A)$. At each \texttt{update} operation, we transfer one of these integers on $T^{j}_b(A)$ (in a cyclic fashion) to avoid overflowing and to achieve worst-case performance. 
Note that each  packed integer is increased by at most $\Delta$ for at most $b-1$ times before being transferred to $T^{j}_b(A)$, so we get the constraint $(b-1)\log ((b-1)\Delta) \leq w$. We choose $(b-1) = \frac{w}{2(\log w + \delta)}$. Then, it is easy to show that the above constraint is satisfied. The number of levels becomes $\log_bn  = \bigO(\log_{w/\delta} n)$. Since we spend constant time per level, this is also the worst-case time needed to answer \texttt{sum} and \texttt{update} queries on our structure.
To analyze space usage we use the corrected formula 
$$
\begin{array}{l}
S_b(1,k) = k\\
S_b(n,k) = \frac{n(b-1)}{b}\cdot \left(k+\log b\right) + \frac{nw}{b} + S_b(n/b,k+\log b)
\end{array}
$$
yielding $S_b(1,k) \leq nk + 2n\log b + \frac{nw}{b-1}$. Replacing $b-1 = \frac{w}{2(\log w + \delta)}$ we achieve $nk + \bigO(n\delta + n\log w)$ bits of space.

We now apply the sampling technique of Section \ref{sec:sampling} with a slight variation. In order to get the claimed space/time bounds, we need to further apply bit-parallelism techniques on the packed integers stored in $A$: using techniques from~\cite{hagerup1998sorting}, we can answer  \texttt{sum}, \texttt{search}, and \texttt{update} queries in $\bigO(1)$ time on blocks of $w/k$ integers.  It follows that we can now use sample rate $d = \frac{w\log n}{k\log(w/\delta)}$ without affecting query times. After sampling $A$ and building the Fenwick tree above described over the sums of size-$d$ blocks of $A$, the overall space is $nk(1-1/d) + S_b(n/d, k+\log d) = nk + \frac{n\log d}{d} + \bigO(\frac{n\delta}{d}+ \frac{n\log w}{d})$. Note that $d \leq \frac{w^2}{k\log(w/\delta)} \leq w^2$, so $\log d \in\bigO(\log w)$ and space simplifies to $nk + \bigO(\frac{n\delta}{d}+ \frac{n\log w}{d})$. The term $\frac{n\delta}{d}$ equals $\frac{n\delta k \log(w/\delta)}{w\log n}$. Since $\delta \leq w$, then $\delta\log(w/\delta)\leq w$, and this term therefore simplifies to $\frac{nk}{\log n} \in o(nk)$. Finally, the term $\frac{n\log w}{d}$ equals $\frac{n\log w \cdot k \log(w/\delta)}{w\log n} \leq \frac{nk}{(w\log n)/(\log w)^2} \in o(nk)$. The bounds of Theorem \ref{th2} follow.

\subsubsection{Parallelism} Note that \texttt{sum} and \texttt{update} queries on our succinct Fenwick trees can be naturally parallelized as all accesses/updates on the levels can be performed independently from each other. For sum, we need $\bigO(\log\log_bn)$ further time to perform a parallel sum of the $\log_bn$ partial results. It is not hard to show that---on architectures with $\log_bn$ processors---this reduces \texttt{sum}/\texttt{update} times to $\bigO(\log\log_bn)$/$\bigO(b)$ and $\bigO(\log\log_{w/\delta} n)$/$\bigO(1)$ in Theorems \ref{th1} and \ref{th2}, respectively.

%
%
%
